\documentstyle[aps,prl]{revtex}
\begin{document}
\def\bra#1{\mathinner{\langle{#1}|}}
\def\ket#1{\mathinner{|{#1}\rangle}}
\def\aver#1{\mathinner{\langle{#1}\rangle}}
\newcommand{\zzop}{\renewcommand{\arraystretch}{0}
\begin{array}c \circ\\ \mid \\
\circ\end{array}\renewcommand{\arraystretch}{1}}
\include{psfig}
\draft
\wideabs{
\title{\bf
Josephson junction qubit network with current-controlled interaction }
\author{J. Lantz, M. Wallquist, V.S. Shumeiko, and G. Wendin}
\address{Department of Microtechnology and Nanoscience, Applied Quantum
Physics Laboratory\\
Chalmers University of Technology, SE-41296 Gothenburg, Sweden.\\}

\date{\today}
\maketitle
\begin{abstract}
We design and evaluate a scalable charge qubit chain network with controllable
current-current coupling of neighbouring qubit loops via local dc-current gates.
The network allows construction of general N-qubit gates. The proposed design is
in line with current main stream experiments
\end{abstract}
}

\narrowtext

Although a working solid-state quantum computer with hundreds of
qubits remains a distant goal, coupling of a few solid-state qubits is now
becoming feasible. Several groups have succeeded to operate single Josephson
junction (JJ) qubits
\cite{Nakamura,Chiorescu,Vion,Yu,Martinis,Ilichev,Duty,Buisson}, but the art
of multiple JJ qubit gates is still in its infancy. A few challenging
experiments with coupled JJ-qubits have been reported
\cite{Pashkin,Berkley,Majer,Yamamoto,Izmalkov}.
However, so far experiments on coupled JJ qubits have been performed
without direct physical control of the qubit-qubit coupling.

There are many proposed schemes for  two(multi)-qubit gates where an
effective qubit coupling is controlled by tunings of qubits or bus resonators
\cite{Shnirman,Makhlin,Plastina,Blais}. However, there are also suggestions
how to control physical qubit interaction
\cite{Averin,Siewert,YouPRB,YouPRL,AverinBruder}, most of which require local
magnetic field control. Recently, Yamamoto {\em et al.} \cite{Yamamoto}
successfully implemented a CNOT gate using fixed capacitive coupling between
two charge qubits, controlling the effective qubit-qubit interaction by
tuning single-qubit level splittings into resonance - however, this method
might not be well suited for more advanced gates on charge qubits because of strong decoherence when qubits are operated away from the degeneracy
points.

In this paper we present a new solution for controllable physical
qubit-qubit coupling, as shown in Fig. 1. The network has the following
properties:
(a) nearest-neighbour qubit-qubit coupling controlled by external bias
current,
(b) qubits parked at the degeneracy points, also during qubit-qubit interaction,
(c) separate knobs for controlling individual qubits and qubit-qubit coupling,
(d) scalability.
An important feature is that the network is easily fabricated, and is in line
with current mainstream experiments.

The network under consideration consists of a chain of charge qubits - Single
Cooper Pair Transistors (SCPT) - with loop-shaped electrodes coupled together
by current biased coupling JJs at the loop intersections (Fig.\ 1).
The loop-shaped electrode was introduced \cite{Makhlin,Nakamura} to provide
external control of the Josephson coupling of the qubit island to the reservoir.
The loop design creates an (inductive) interface to the qubit by means of
circulating currents \cite{Zorin}, which has been used as a tool for qubit
readout by Vion et al.\cite{Vion}. We employ these current states in the qubit
loops to create controllable coupling of neighbouring qubits.
The results of this paper are derived in the charge qubit limit $E_C\gg E_J$.
However, the analysis and the coupling mechanism also apply to the case of
$E_C\approx E_J$, describing the charge-phase qubit \cite{Vion}.

\begin{figure}[t]
\centerline{\psfig{figure=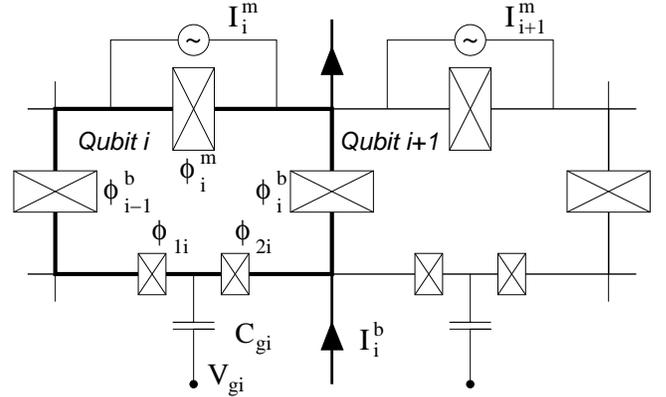,width=8.5cm}}
\centerline{\caption{
Network of loop-shaped SCPT charge qubits, coupled by large Josephson junctions.
The interaction of the qubits ($i$) and ($i+1$) is controlled by the current bias $I^b_i$
or by simultaneous current biasing of readout junctions.
Individual qubits are controlled by voltage gates, $V_{gi}$.
Single-qubit readout is performed by applying an ac
$[25,26]$ or dc $[3]$
current pulse $I^m_i$ to a particular JJ readout junction.
Alternatively, readout of charge may be performed, eg.\ using a RF-SET capacitively
coupled to the island $[27]$
The two Josephson junctions of the $i$-th SCPT are assumed to have identical Josephson
energies $E_{Ji}$. The Josephson energies of the coupling JJs $E_{J}^b$ and the readout
JJs $E_{J}^m$ (identical for simplicity) are much larger than the corresponding
charging energies, $E_{J}^b \gg E_{C}^b$. $\phi^b_i$ ($\phi^m_i$) is the phase
difference across the $i$-th coupling (readout) JJ. For an open N-qubit chain we
choose $\phi^b_{0}=\phi^b_{N}=0$, while for a closed chain $\phi^b_{0}=\phi^b_{N}$.}}
\end{figure}

Left without any external current biasing of the coupling and readout JJs,
the network acts as a quantum memory of independent qubits (neglecting a
weak residual interaction, to be discussed below). When a bias current is sent through
the coupling JJ in Fig.1, the current-current interaction between the neighbouring qubits
is switched on and increases with increasing bias current. Moreover, if {\em both} of the
readout JJs of the same qubits are biased well below
threshold, again there is nearest-neighbour coupling via the circulating
currents \cite{YouComment}. However, if {\em one} of the readout JJs is current biased,
this only affects that particular qubit and allows readout of individual qubits.
Thus the bias currents serve as the interaction control knobs.
The loop inductances are assumed to be sufficiently small to neglect
qubit-qubit coupling via induced magnetic flux, as well as undesirable
qubit coupling to the magnetic environment. In addition, we assume
negligible capacitive coupling between the islands, which are well
shielded by the injection leads.

The SCPT qubit chain system shown in Fig.\ 1 is described by the
Hamiltonian
\begin{equation}
{\cal H}=\sum_{i} (H_{\mbox{\tiny SCPT},i} + H_{\mbox{\tiny OSC},i}^b+
H_{\mbox{\tiny OSC},i}^m),
\label{H1}
\end{equation}
where $H_{\mbox{\tiny SCPT},i}=(E_{Ci}/2)(1-n_{gi}) \sigma_{zi} -
E_{Ji}\cos\theta_i \sigma_{xi}$, and where
$H_{\mbox{\tiny OSC},i}^b= Q_i^2/2C^b-E_{Ji}^b\cos\phi_i^b - {\hbar\over
2e}I_i^b\phi_i^b$ is the Hamiltonian of the coupling JJ.
$H_{\mbox{\tiny OSC},i}^m$ is the similar Hamiltonian of the readout
junction, which for simplicity is chosen with the same parameters.
$en_{gi}=C_{gi}V_{gi}$ is the induced gate charge on the $i$-th SCPT
island, the charging energy of which is defined by $E_{Ci}=(2e)^2/2C_{\Sigma}$.
Finally, $Q_i$ is the charge on each coupling JJ obeying the
commutation relations $[Q_{i},\phi^b_{j}]=2ie\delta_{ij}$.
$H_{\mbox{\tiny SCPT},i}$ has been truncated to the two lowest charge
states, assuming $E_{Ci}\gg E_{Ji}$ and $n_{gi}\approx 1$, and a small correction
to the charging energy of the coupling
JJs ($\sim C_i/C^b \ll 1$) has been neglected.
The flux quantization $\phi_{i1}+\phi_{i2}-2\theta_i = 0$, in
every loop (assuming zero external flux) introduces a dependence of the
qubit Josephson energy on the phase differences across the coupling
and readout  JJs,
$\theta_i=(\phi_i^{b}-\phi_{i-1}^{b}-\phi_i^m)/2$. This qubit-oscillator
interaction is the origin of the qubit-qubit interaction.

For proper functioning of the network, the critical conditions are
$E_J^b \gg \hbar\omega_p, E_J$, where $\omega_p$ is the plasma frequency
of the coupling JJs, establishing the phase regime for the coupling JJs
with small fluctuations of phase
$\delta_i=\phi_i^{b}-\bar{\phi}_{i}^{b}$, $<\delta_i^2> \;\sim \;
\hbar\omega_p/E_J^b \ll 1$,
around energy minima determined by the control current,
$\sin\bar{\phi}_i^b=I_i^b/I_c^b$.
We only consider the regime of negligible macroscopic quantum tunneling
(MQT).

Using a harmonic approximation for the periodic potential terms in Eq.\
(\ref{H1}), all coupling JJs are reduced to harmonic oscillators with
level spacing $\hbar\omega_p=\sqrt{2\epsilon^b_iE_C^b}$, where
$\epsilon_i^b=E_{J}^b\cos\bar{\phi}_i^b$. Each SCPT term in Eq.\ (1) is
then, in the lowest approximation with respect to harmonic amplitudes,
reduced to a qubit Hamiltonian,
\begin{equation}
H_{qi}={E_{Ci}\over 2} (1-n_{gi}) \, \sigma_{zi} -
E_{Ji}\cos\bar{\theta}_i \,\sigma_{xi}, \label{Hq}
\end{equation}
plus a linear oscillator-qubit interaction, $H_{\mbox{\tiny
int},i}^{(1)}=\lambda_i (\delta_i-\delta_{i-1})\sigma_{xi}$,
proportional to the
coupling strength $\lambda_i=E_{Ji}\sin\bar{\theta}_i$
and to the
phase deviation in the coupling JJs. This generates
controllable nearest-neighbour qubit interaction terms which appear only
in the presence of bias currents and describe displacement of the
oscillators driven by the qubits. There are also quadratic terms,
$H_{\mbox{\tiny int},i}^{(2)} =
(\epsilon_i/2)(\delta_i-\delta_{i-1})^2\sigma_{xi}$, where
$\epsilon_i=E_{Ji}\cos\bar{\theta}_i$, which induce relatively
small permanent residual qubit coupling due to oscillator squeezing
driven by the qubits.

The harmonic oscillators can with good accuracy be assumed to stay in
the ground state during all quantum operations on the network at low
temperature.
For current control pulse durations $T \gg \omega_p^{-1}$, the
probability to excite the oscillator due to qubit flips away from the
degeneracy point is estimated as $W\;\sim
E_{Ci}^2(1-n_{gi})^2/\hbar\omega_{p} \epsilon_J^b \ll 1$, when the
linear qubit-oscillator coupling is switched on ($\lambda_i \sim
E_{Ji}$). In the residual interaction regime ($\lambda_i = 0$), the
excitation probability is several orders of magnitude
($\hbar\omega_{p}/E_J^b \ll 1$) smaller.
Hence, we average over the ground state of the oscillators and finally
arrive at the effective Hamiltonian for the qubit network,
\begin{equation}
{\cal H}= \sum_i(H_{qi} +\eta_i\,\sigma_{xi}\sigma_{x(i+1)}) +
\sum_{i\neq j}\kappa_{ij}\,\sigma_{xi}\sigma_{xj}
\label{H2}
\end{equation}
where
$\eta_i=\lambda_i\lambda_{i+1}/\epsilon_i^b$ and $\kappa_{ij}$ are the
energies of the controllable and the residual interactions,
respectively. The maximum controllable interaction energy is a factor of
$E_J/\epsilon_i^b$ smaller than the qubit level splitting, $\sim 2E_J$.
The residual qubit-qubit interaction effectively connects all of the
qubits but it is smaller than the controllable coupling by a factor of
$\hbar\omega_p/E_J^b \ll 1$.

The interaction energy $\eta_i$ can be expressed in terms of the
currents $I_{i}=(e/\hbar)E_{Ji}\sin\bar{\theta}_i\label{i}$
circulating in neighbouring qubit loops as $\eta_i=L_{\mbox{\tiny eff}}
I_{i}I_{i+1}$, where $L_{\mbox{\tiny eff}}=\hbar^2/(4e^2\epsilon^b_i)$
is the effective inductance introduced by the coupling JJ.

In order to exclusively couple the qubits ($i$) and ($i+1$) one should
apply a nonzero current bias $I^b_{i}$, while $I^b_{i\pm 1}=0$ and
$I^b_{i\pm 2}=0$. In this case the coupling amplitude is given by
the equation,
\begin{equation}
\eta_i=-{E_{Ji}E_{J(i+1)}\over
4E_{J}^b\cos\bar{\phi}_i^b}\sin^2{\bar{\phi}_i^b\over 2}.
\label{eta}
\end{equation}
The coupling is quadratic, $\sim (I^b_i/I_c^b)^2$, for small current
bias, and diverges when approaching the critical current. An alternative way
to switch on the qubit-qubit coupling is to apply dc bias currents (below the
critical value) simultaneously to both of the two neighboring readout junctions,
instead of activating the coupling junction, resulting in an interaction energy
$\eta_i=-(E_{Ji}E_{J(i+1)}/4E_{J}^b) \sin(\bar{\theta}_i/2) \sin(\bar{\theta}_{i-1}/2)$.

The present strategy is to park the qubits at the degeneracy points, where the
coherence time is maximum \cite{Vion}, and then to operate with (a) short
dc-voltage pulses or, alternatively, microwave resonant excitation, to perform
single-qubit operations with qubit-qubit coupling switched off ($\eta=0$), and
with (b) dc-current pulses ($\eta \ne 0$) to perform two-qubit rotations at the
degeneracy points.

The readout of individual qubits can be performed by probing the corresponding
junction with ac current \cite{Siddiqi1,Siddiqi2}, while keeping zero bias at
the coupling junctions. Since amplitude of the phase oscillation in the readout
junction remains small during measurement, our theory applies, and neighboring
qubits will not be disturbed. Another option would be to pulse the dc current
through the readout junction above the critical value \cite{Vion}. Required
that the qubits are operated in the charge regime ($E_C\gg E_J$), readout of
the islands charge state is also possible by means of a capacitive probe, eg.
using a SET electrometer.

We now focus on two-qubit gate operation.
The Hamiltonian in Eq.\ (\ref{H2}) is diagonal in the current basis when
all qubits are put at the degeneracy points.
Considering two neighbouring qubits, 1 and 2 in the current eigenbasis
$\ket{00},\ket{01},\ket{10},\ket{11}$, the Hamiltonian explicitly
becomes (assuming for simplicity identical qubits $\epsilon=\epsilon_1=\epsilon_2$),
\begin{equation}
{\cal H}=\left(\begin{array}{cccc} 2\epsilon+\eta_1 &  0 &  0 &  0 \\
  0 &  -\eta_1 & 0 &  0 \\
  0 &  0 &  -\eta_1 &  0 \\
  0 &  0 &  0 & -2\epsilon+\eta_1 \end{array}\right),
\end{equation}
where $\eta_1$ is given by Eq.\ (\ref{eta}),
We can now use the current control bias to perform coupled-qubit phase
rotations \cite{Plastina}. We define a basic entangling two-qubit gate operation, the
$-\pi/2$ zz-rotation,
\begin{equation}
\zzop\equiv e^{-{i\over\hbar}\int_T\,dt{\cal
H}(t)}\sim\left(\begin{array}{cccc}  i & 0 & 0 & 0 \\
0 & 1 & 0 & 0 \\
0 & 0 & 1 & 0 \\
0 & 0 & 0 & i \end{array}\right)
\end{equation}
by choosing the appropriate amplitude $I^b$ (i.e. $\eta_1$)  and
duration $T$ of the bias current pulse, determined by the simple
integral equations $\int_T (2\epsilon/\hbar)\,dt=0\mbox{ (mod }\pi)$ and
$\int_T (\eta_1/\hbar)\,dt=-\pi/4$. The operation is only slightly more
complicated for non-identical qubits. The current pulse shape is of no
importance,
except that it must be adiabatic with respect to the harmonic degrees of
freedom, $\eta_i\ll\hbar\omega_p$.

By means of the $-\pi/2$ zz-rotation and single qubit rotations it is
now
straightforward to construct any desired quantum operations, including
generalized quantum gates. Standard two-qubit gates such as the CNOT
operation  require a short sequence of additional single qubit $\pi/2$
rotations \\

\begin{figure}[t]
\centerline{\psfig{figure=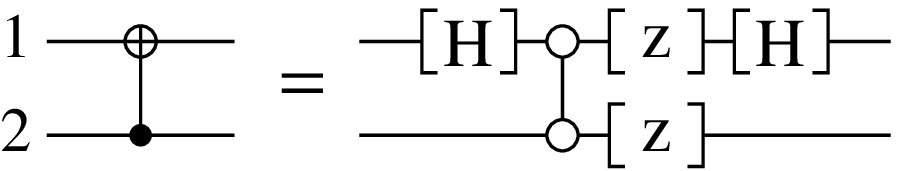,height=1cm}}
\end{figure}
\noindent
where $[k]_i=\exp[-i(\pi/4) \tau_{ki}]$, $\tau_{ki}$ are Pauli matrices
in the current basis, and the Hadamard operation [H] corresponds to the
sequence $[$x$][$z$][$x$]$.
Another useful operation, CNOT-SWAP, can be also introduced, \\

\begin{figure}[t]
\centerline{\psfig{figure=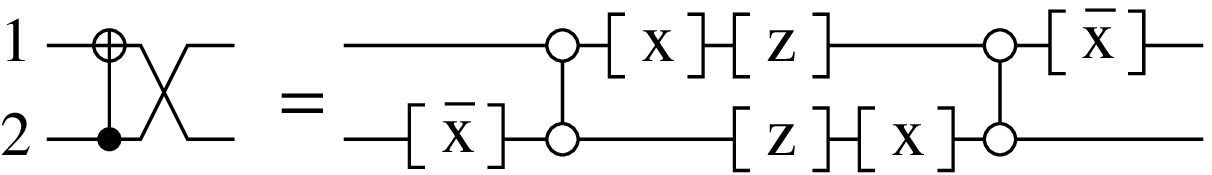,height=1cm}}
\end{figure}
\noindent
which allows effective implementation of quantum algorithms on qubit
networks with nearest neighbour interaction \cite{Schuch1,Schuch2}. The
operations have been optimized in the sense that the $[z]$ rotations can
be performed using the natural precession of the qubits.

\begin{figure}[t]
\centerline{\psfig{figure=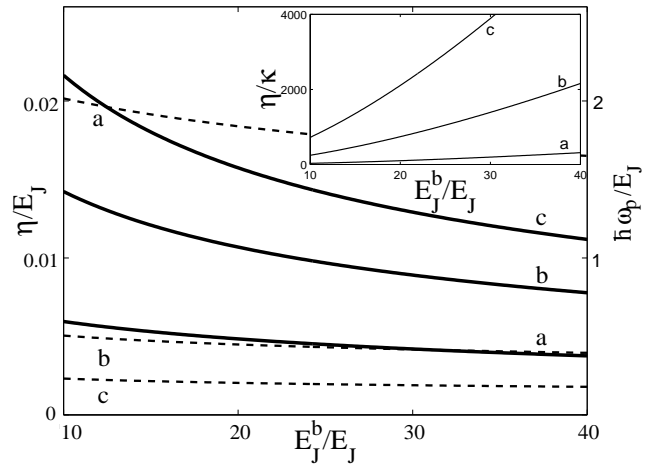,width=8.5cm}}
\centerline{\caption{ Maximum coupling energy $\eta$ of two neighbouring
qubits with a current bias applied to the coupling JJ which corresponds to
the MQT rate $\Gamma_{\mbox{\tiny MQT}}=(2E_J/h)\times 10^{-4}$, for
different shunt capacitances (solid), $C_{shunt}^b=0\;(a)$, $10C^b\;(b)$,
$40C^b\;(c)$. The corresponding plasma frequency, for the same parameters,
(dashed) decreases with increased $C_{shunt}^b$. Inset: the ratio of the
controllable interaction and the residual interaction.}}
\end{figure}

The time needed for a two qubit operation is given by the coupling strength
$\eta$, whose upper limit is set by MQT in coupling JJs, and depends on the
plasma frequency: a lower plasma frequency yields lower rate of MQT and thus
higher maximum current bias. Thus, a stronger controllable coupling is
achieved by adding a large shunt capacitance $C^b_{shunt}$ to reduce the
plasma frequency. On the other hand, the latter is restricted by the
adiabaticity condition, $\eta_i\ll\hbar\omega_p$. 
The maximum coupling energy can be estimated using the standard
expression describing MQT in current biased JJs \cite{Weiss}, neglecting the
small circulating currents, $\Gamma_{\mbox{\tiny MQT},i}\approx
\bar\omega_p(60 s/\pi)^{1/2}\exp(-s)$, where
$s=(24E_{J}^b/5\hbar\bar\omega_p)(\cos\bar\theta_i)^{5/2}$ 
and $\bar\omega_p = (2E_J^b \bar E_C^b)^{1/2}$ 
is the bare plasma frequency of the 
capacitively shunted coupling junction, $\bar E_C^b =
(2e)^2/2(C^b+C^b_{shunt})$.
Requiring the lower bound for $\cos\bar\theta_i$ to be larger than
$(\bar\omega_p/E_J^b)^{2/5}$ under condition that the MQT rate remains
negligibly small, the adiabaticity condition gives,
$\bar\omega_p \gg E_J(E_J/E_J^b)^{1/4}$, and the maximum coupling energy in
Eq. (\ref{eta}) becomes $\eta_{max} \sim E_J (E_J/E_J^b)^{1/2}$. 
Quantitative results for the maximum coupling energy are shown in Fig. 2. Note, that the residual interaction is reduced for small plasma frequency (see inset in Fig. 2).
Taking the interaction energy to be $\eta_{max} =(2E_J)/100$, the time needed for the $-\pi/2$ zz-rotation is then approximately 25 times the qubit period time ($h/2E_{J}$).

Assuming that the qubits are operated at the degeneracy point,
fluctuations in the biasing current will cause pure dephasing. Nevertheless, the
qubits will be decoupled from current noise to first order at zero
current bias. However, since relatively long periods of qubit coupling
are needed to perform practical control operations, suppression of bias
current fluctuations might be essential \cite{Martinis}.

Finally it should be emphasized that although this paper has been concerned with
the charge qubit limit $E_J/E_C \ll 1$, the analysis and the design for
bias-current-controlled qubit-qubit coupling is equally relevant in the
region of $E_J/E_C \approx 1$, characterizing the "Quantronium" charge-phase
qubit \cite{Vion}.
A higher $E_J/E_C$ ratio introduces more charge states and flattens the bands,
making the system less sensitive to background charge fluctuations.
The coupling of neighbouring qubits will however still be controllable, and
higher levels will not be excited during two-qubit gate operations provided
that the bias current is switched on adiabatically.

In conclusion, the present scheme provides a realistic solution for easy local control
of the physical coupling of charge qubits via current biasing of
coupling Josephson junctions or, alternatively, pairs of readout
junctions. The design is in line with experimental mainstream
development of charge qubit circuits and can easily be fabricated and
tested experimentally. Most importantly, it allows readout via currently
tested methods that promise single-shot projective measurement and even
non-destructive measurements, via e.g. RF-reflection readout of a JJ threshold
detector \cite{Siddiqi1,Siddiqi2} or an SET \cite{Duty,Aassime}.
The tunable coupling of the qubit chain allows easy implementation of
CNOT and CNOT-SWAP operations. Independent two-qubit operations can be
performed in parallel when the network consists of five qubits or more,
and generalization to single-shot N-qubit gates seems possible.
This may offer interesting new
opportunities for operating qubit clusters in parallel and swapping
and teleporting qubits along the chain, for experimental implementations of
elementary quantum information processing
\cite{Siewert,Schuch1,Gingrich,Schuch2,Brennen,AverinFazio}.

Acknowledgment - We thank Y. Nakamura, D.Esteve, R. Fazio and P. Delsing for helpful discussions. This work has been supported by the EU IST-FET-SQUBIT project,
the Swedish Research Council, and the Swedish Foundation for Strategic Research.


\begin{thebibliography}{999}
\bibitem{Comment1}
The loop-shaped electrode was introduced \cite {Makhlin,Nakamura} to
provide external control of the Josephson coupling of the qubit island
to the reservoir.
\bibitem{YouComment}
For two qubits, similar controllable coupling can be achieved by
applying flux bias \cite{YouPRB} instead of current bias. However, to be
scalable this approach requires complicated circuitry \cite{YouPRL}.
\end{thebibliography}
\end{document}